\documentclass[conference]{vgtc}                     

\onlineid{0}

\vgtccategory{Research}
\vgtcpapertype{Application/Design Study}                  

\graphicspath{{figures/}{pictures/}{images/}{./}} 
\usepackage{mathptmx}                  

\title{A Lens to Pandemic Stay at Home Attitudes}

\author{\authororcid{Andrew Wentzel}{0000-0002-2003-2750}, Lauren Levine, Vipul Dhariwal, Zahra Fatemi, \authororcid{Abari Bhattacharya}{0000-0003-3647-3363},\\ \authororcid{Barbara Di Eugenio}{0000-0003-1706-2577},  \authororcid{Andrew Rojecki}{0000-0003-0773-9011}, \authororcid{Elena Zheleva}{0000-0001-7662-2568}, \authororcid{G.Elisabeta Marai}{0000-0002-7212-9669}}

\authorfooter{
A. Wentzel, L. Levine, V. Dhariwal, Z. Fatemi, A. Bhattacharya, B. Di Eugenio, A. Rojecki, E. Zheleva, and G.E. Marai are with the University of Illinois Chicago. E-mail: {awentze2@uic.edu}.\\
}

\teaser{
    \includegraphics[width=\textwidth]{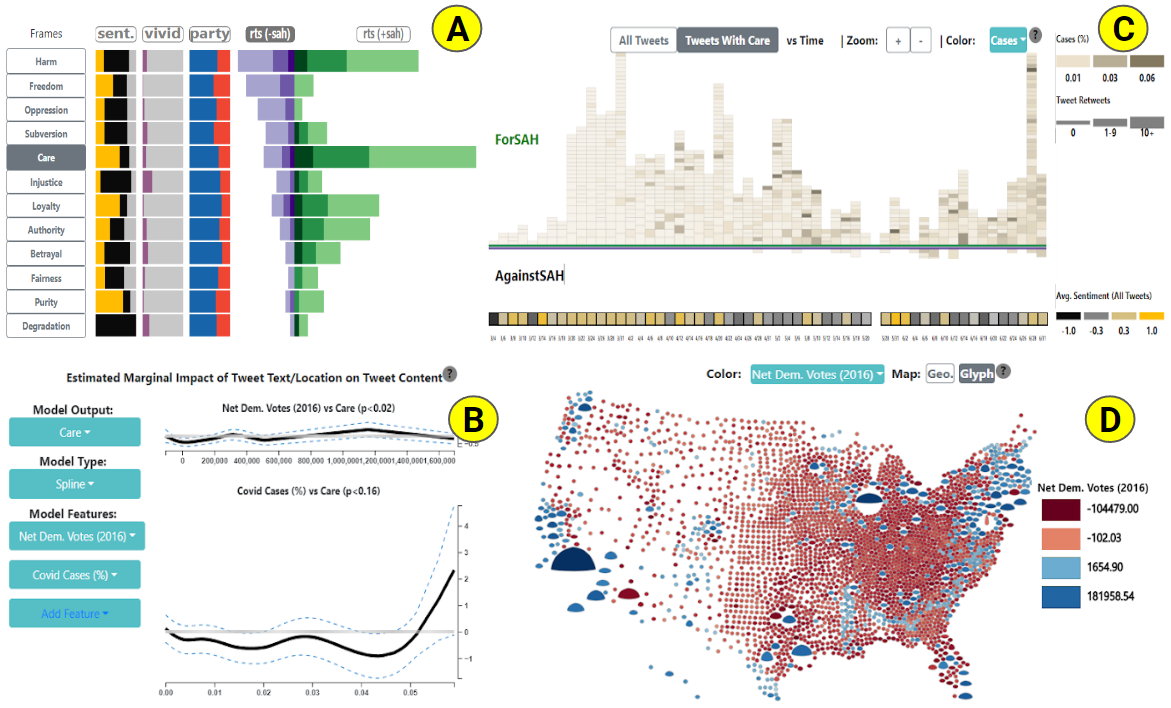}
    \caption{Analysis of moral frames (MF) in social media posts related to Stay-at-Home orders during the COVID-19 pandemic. (A) Tweet summary showing several MF tweet features, sorted here by tweets with a negative stance. (B) Inference panel showing partial dependence plots derives from generalized linear models (GAMS), here showing the relationship between county voting history and COVID-19 cases, and Care tweets. (C) Timeline of Care tweets, along with popularity, COVID-19 incidence, and sentiment. Note the negative spikes in April/June, after BLM protests, and the more negative sentiment (lower bar). A tooltip shows the text of a tweet supporting SAH orders. (D) Glyph-map of counties showing political party (color) and population (width) vs. Care tweets (height); major cities and rural white areas stand out.}
    \label{fig:teaser}
}

\abstract{
We describe the design process and the challenges we met during a rapid multi-disciplinary pandemic project related to stay-at-home orders and social media moral frames. Unlike our typical design experience, we had to handle a steeper learning curve, emerging and continually changing datasets, as well as under-specified design requirements, persistent low visual literacy, and an extremely fast turnaround for new data ingestion, prototyping, testing and deployment. We describe the lessons learned through this experience.
}

\begin{document}

\maketitle

\section{Introduction}
In early 2020, in light of the emergence of the global COVID-19 pandemic, the U.S. National Science Foundation (NSF) began accepting proposals for non-medical, non-clinical research that could be used right away to explore how to model and understand the spread of COVID-19, how to inform and educate about the science of virus transmission and prevention, and how to encourage the development of procedures and actions to address the pandemic~\cite{NSFcall}. NSF invited researchers to apply for funding through the Rapid Response Research (RAPID) mechanism. This mechanism enables the NSF to receive and review proposals that have a high priority in terms of the availability of or access to data, facilities, or specialized equipment, as well as quick-response research on natural or anthropogenic disasters and similar unforeseen events. 
Requests for RAPID proposals could be for up to \$200K and up to one year in duration. 

Our collaborative idea was for a RAPID project analyzing people’s moral take on and consequent response to government mandated stay-at-home (SAH) orders at the start of the pandemic, leading to potential insights into effective governmental messaging of such orders. Shared values, beliefs, and understandings were often reflected at the time in social media posts, therefore, we were interested in identifying the underlying values expressed in such posts, and how they related to attitudes with respect to SAH messaging. Concretely, the funded project aimed to analyze these values and attitudes using Moral Foundations (MF) theory, a psychological model for describing the different dimensions underlying social discourse, 
 such as Care or Liberty or Loyalty. Building from the strengths of the team, the project focused on applying MF theory, in collaboration with social-science researchers, Causal Inference (CI) researchers, and Natural Language Processing (NLP) researchers to help build and analyze a MF-annotated tweet corpus related to SAH orders. Additionally, the project aimed to leverage geospatial data to analyze datasets at multiple levels of spatial aggregation, and to compare temporal and spatial differences to enhance participation and promote positive public health outcomes.

Analyzing social media from a moral-framing perspective posed a number of challenges. First, while collaborators had clear insights into MF theory, applying these frames to social media was a difficult and poorly defined task. 
an example, a tweet stating “We are not staying home!” could be an expression of the “Liberty” frame in opposition to government lockdowns if it was tweeted at the end of March 2020. However, if the tweet occurred two months later during a major civil movement protest, it could be a tweet about solidarity with other protesters and thus would be an expression of the “Loyalty” frame. 
Therefore, meaningful analysis of the data requires understanding the tweeter's situational context, such as time, location, and major recent events. Second, since developing a corpus to analyze was part of the proposed work, design requirements had to constantly be updated as the expected scope of the data and resulting expectations changed week-to-week. We also found that annotating tweets with moral frames is a difficult task that requires trained experts. These issues resulted in rapid changes in the expected number of tweets that were both annotated and geotagged, the ability to tie multiple tweets to individual users, and the kinds of textual features that were deemed relevant. This resulted in us moving to an agile design approach, with rapid prototyping cycles that constantly changed. Finally, there were many design challenges that were imposed by the collaboration itself. For example, we found that our collaborators, who had a diverse range of backgrounds and experience levels, tended to have limited visual literacy. While the resulting design process was difficult, the collaboration did result in usable insights and results, with results being published at conferences~\cite{APSApaper,elenaDSAA}.

\section{Related Work and Background}
Our team leveraged Moral Foundation Theory (MFT) for this project. MFT is a psychological model for describing the different dimensions underlying social discourse. MFT is widely used to study how values differ between individuals, and has been used to explain differences in political affiliation and belief systems~\cite{koleva2012tracing}. Our version of MFT considered 6 foundation frames with opposing pairwise "virtues" and "vices" qualities, such as "Care" (virtue), which pertains to "the need to help or protect oneself or others", and the corresponding "Harm" (vice), which deals with "fear of damage or destruction to oneself or others"~\cite{graham2013moral}.

Recent visual analysis studies had looked at attitudes on social media related to public health, including attitudes towards the coronavirus~\cite{aiello2021epidemic}, public health interventions~\cite{doogan2020public,hu2021revealing}, climate change~\cite{diakopoulos2014identifying}, and popular topics of discourse on social media~\cite{jang2021tracking,chang2021people,chandrasekaran2020topics}. Other systems have looked at information spread on social media,  journalism, and misinformation~\cite{googleripples,rodgers2008general,mishra2022news,dant2011behind,karduni2019vulnerable}, or focused on real-time information spread without geolocation~\cite{whisper,rmap,knittel2021real}. Systems that focus on event detection~\cite{leadline,scatterblogs2} tend to not analyze stance or moral foundations. Systems have also focused on temporal progression of topics~\cite{sententree, digitalbackchannels,opinionflow,cobridges,kucher2020stancevis,fluxflow,targetvue}, but none of these systems tie discourse to demographics or MFT framing.

\section{Methods}
\begin{figure}[htbp]
    \centering
    \includegraphics[width=\linewidth]{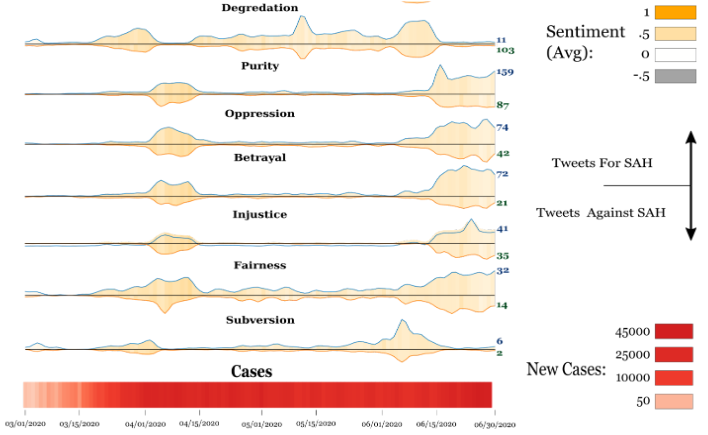}
    \caption{ Design of timelines using aggregate tweets and small multiples for each tweets for each moral frame used on a larger non-geotagged dataset.}
    \label{fig:timeline}
\end{figure}

\begin{figure*}[htbp]
\centering
    \includegraphics[width=0.8\linewidth]{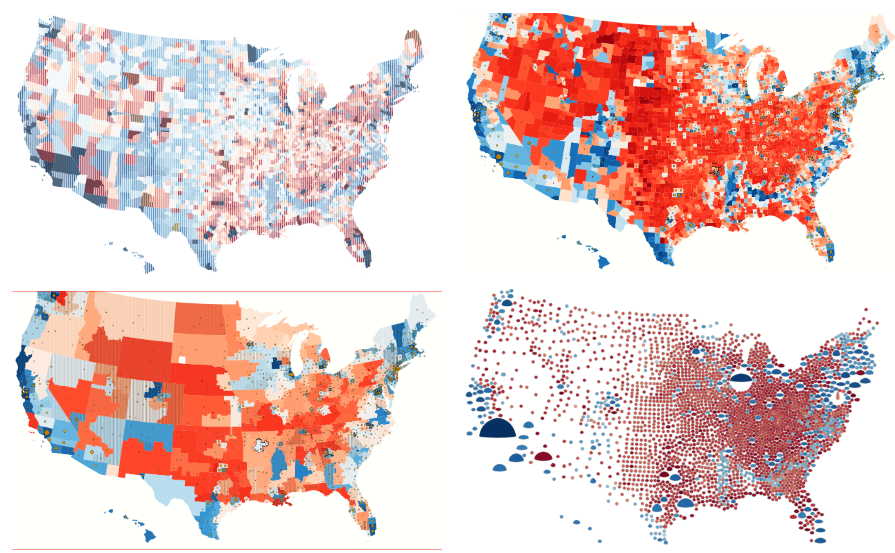}
\caption{Examples of the progression of map design through the design process. (Top-left) Choropleth map using texture weaving~\cite{hagh2007weaving} to encode two features. (Top-right) Choropleth map with a single color and glyphs showing tweet distribution. (Bottom-left) Choropleth map using textures and glyphs. Areas are aggregated by the intersection of district voting maps and counties for demographic data to make each section more evenly distributed in terms of population. Glyphs show tweets aggregated at the county level. (Bottom-right) Glyph-based map where shape encodes tweet features and color encodes demographics. All maps are zoomable}
    \label{fig:maps}
\end{figure*}

Our group had to rely on remote collaboration due to the pandemic lockdowns, with weekly meetings, and extensive use of collaborative tools leveraging our laboratory's expertise~\cite{marai2019immersive}. The core group spanned four research labs, evenly distributed: communications, NLP, causal inference in social media, and visual computing. Per our initial agreement, all group members are listed as co-authors. Our design process relied heavily on rapid prototyping using an activity-centered approach~\cite{marai2017activity} and an agile framework, a combination which had been very successful in our work with domain experts across the disciplines~\cite{maries2013grace, maries2012interactive,luciani2019details, wentzel2019cohort}. The project featured two stages, foraging for data and features, and hypothesis testing, which focused on establishing the cause of trends. Our collaborators’ research activities were focused on identifying and explaining SAH-related phenomena and how Moral Framing factors into them. In short, we used MF as a lens to create a SAH-relevant dataset, and then to analyze and describe larger social trends. This agile, distributed design was similar to the distributed prototyping methods described by Losev et al~\cite{losev2020distributed}. However, our process focused more heavily on rapid prototypes without interaction design due to a larger emphasis on the data foraging stage.

\subsection{Overall Design and Data Sources}
The overall top-design emerged gradually and organically, and featured multiple coordinated views (\cref{fig:teaser}). We gradually added data sources to our system, starting with a continually updated and expanded set of manually labeled geotagged U.S. tweets that expressed a moral frame and stance regarding stay-at-home (SAH) orders between March and May 2020. Tweets were manually labelled for SAH relevance, moral framing, and stance by expert annotators. Non-relevant tweets or tweets with no moral frames were excluded. The resulting corpus is described by Fatemi et al.~\cite{elenaDSAA}. During processing, each tweet was mapped to one of the 3113 U.S. counties, not including Alaska as we did not have election data for those areas. We then integrated several geospatial datasets: 2018 census information~\cite{medsl2018elections,census2018}, voting results from the 2016 US presidential election results~\cite{medsl2018elections}, results from a New York Times survey on estimated mask usage~\cite{nyt2020mask}, and daily COVID-19 cases and deaths for each county~\cite{johnshopkins2020dashboard}. We also gradually extracted tweet features such as stance (pro SAH or against SAH), virality, sentiment, and vividness, and aggregated each moral frame stance within each county, resulting, over the length of the project, in 14 different continuous values for each county.

\subsection{Four Custom Views}
One view was dedicated to data summarization, and supported foraging activities at the start of the project. It gives an overview of the tweet features of interest, such as sentiment or vividness, broken down by Moral Frame. Due to limited visual literacy among the group at the start of the project, the view made extensive use of rotated, colored bar charts. To alleviate the cognitive load due to the extensive use of color, the colors were mapped as intuitively as possible (e.g., gray for not vivid, purple for vivid) based on perceptual principles and common media interpretation.

A custom Timeline panel (\cref{fig:timeline}) showed the temporal distribution of the tweets, along with changes in COVID-19 rates, overall sentiment, and tweet popularity. The timeline panel grew with the project, with additional features being mapped to it. Again due to limited visual literacy, the timeline used basic cues such as layout, height, and color (hue and saturation) to encode time-varying multi-dimensional data. Because of scalability issues with this rich custom encoding, later applications to larger datasets used sparklines as an alternative.

Later on during the foraging stage, a Geospatial map panel showed the geographic distribution of tweets with a given Moral Frame, overlaid with demographic data. The design of this panel featured several costly trial-and-error design cycles (\cref{fig:maps}), where texture-blending based approaches which worked on other scientific datasets~\cite{hagh2007weaving}, and which seemed acceptable during lo-fi prototyping, backfired within our group in the hi-fi prototyping stage. We gradually developed an alternative, far more successful glyph encoding.

Finally, during the hypothesis testing stage, providing support for inference analysis became important. An Inference panel was created (\cref{fig:teaser}), which allows for building predictive models within the front-end of the system, to support inferences about how different factors influence average tweet features.


\section{Evaluation}
Our system aimed to support the development of insights related to SAH policy application in the U.S. We report here one of the case studies that our team performed over the course of several months, with results published and presented in several venues~\cite{APSApaper,elenaDSAA}.  The case study (Fig.~\ref{fig:teaser}) was performed remotely, with the team piloting the investigation via Zoom meetings, and the lead author operating the front-end of the system accordingly. Collaborators were also given independent access to the front end, which they used for additional analyses between meetings. For brevity, we distill the main insights here:

\textbf{ Major Frames} The dominant frames were Care, followed by Harm, which share the same foundation. Care tweets were overwhelmingly in support of SAH orders, while Harm were more mixed. Tweets with Harm also increased after the first month of Lockdowns, around times when SAH orders were removed, and had lower sentiment overall, which may reflect pandemic fatigue. Care tweets were significantly correlated with areas that self-reported higher mask use. An important implication of this analysis is that Care-targeted messaging of SAH orders helps.

\textbf{Political Polarization} The frames most associated with Democratic areas were Betrayal and Loyalty, which share a foundation, while the libertarian frames Subversion and Freedom were most associated with Republican areas. Betrayal tweets all happened in the start of May, largely in response to unmasked protests, and the mandated ending of SAH orders in major left-leaning cities located within right-leaning states. Notably, no Betrayal tweets came from Chicago, which may be due to the Mayor's continuation of SAH orders. Subversion and Freedom tweets increased after the first renewal of SAH orders in April. An important implication is that higher-granularity political analysis would help with targeted SAH messaging, for example including a Libertarian perspective in addition to Republican or Democrat views.

\textbf{Spatial-Temporal Trends} Tweets regarding SAH orders peaked at the start of the pandemic, with recurrences focused on major cities when SAH orders were renewed or ended. We also noted a large dip at the end of May, likely due to the George Floyd protests occupying the public zeitgeist at this time, with a short increase at the start of June in response to updates in public mask mandates in several cities. An important finding was that rural areas were underrepresented in the social media dataset.

Qualitative feedback from our collaborators across the disciplines was enthusiastic: \textit{“Excellent for data exploration”}, \textit{“Great for investigating anomalies in the data.”}, \textit{“Amazing work, the encodings work well together.“}, \textit{“This interface does very nicely with the data and domain knowledge we've been given.”}, \textit{“It's nice, pleasant to look at. And extremely informative.”} In terms of ratings, components were rated favorably, with two requests for the additional on-demand glyph pictorial explanation. Experts found the system met these goals well: understanding the MF distribution, temporal MF distribution, MF political context, and verifying MF hypotheses. Most also found the system useful in analyzing MF tweet features, the corpus, exploring sentiment and topic relationships, and MF geographical distribution. The Causal inference and Communications specialists found the system very useful for understanding the emerging corpus, whereas some of the NLP experts continued to rely for corpus analysis tasks on their standard approach (LDA, topic clustering).

\section{Discussion}
Our original design for this project was to create a large-scale analysis tool for analyzing networks of twitter stance and how different moral-framing affected the propagation of certain posts. In reality, we found that there were many difficulties with assessing a developing, real time problem while also incorporating a rich set of features in a collaborative setting. Despite this we did manage to create a unique system despite issues related to availability of data and issues that arose from a short-term collaboration with changing goals. Although our project did not seek to advance Moral Framing theory, our results point to a need for careful consideration of how the MF data is collected, for example using more granular political affiliation measures.

\textbf{Generalizability and Scalability} In response to collaborator and reviewer requests, we expanded our system to be usable for other datasets. Specifically, we relied on an earlier dataset of tweets related to the Black Lives Matter Movement~\cite{hoover2020moral}, where we used hashtag content to estimate stance. The resulting dataset was slightly larger and covered a longer time period, with 1901 geotagged tweets spanning 2 years. The main challenge was adapting the timeline to not encode COVID-19 rates in the timeline, and to use textual features instead. Our results suggest that our approach generalizes well, with the main issue being the availability of geotagged data with sufficient quality annotation. In terms of scalability, we mainly faced issues with adapting the timeline to larger tweets. In a separate analysis using a non-geotagged dataset, we relied instead on a design that relied on small multiples, and shown in \cref{fig:timeline}.


Our collaboration was a learning process in producing rapid, evolving designs, leading to several design lessons, beyond visual scaffolding~\cite{marai2015visual} for improved visual literacy:

\begin{enumerate}

	\item Minimize assumptions about the data when designing during the data collection process. This allows us to gradually specialize designs while keeping earlier progress in the event that the data changes. For example, initially, collaborators' main interest was in the relationship between moral framing and political affiliation. However, data availability made collecting multiple relevant tweets from each individual difficult. Thus we had to rely on location to infer demographics for tweets from each region, and reason about moral framing on an aggregate scale. Given the necessity of context, we further found that automatic annotation of moral frames performed too poorly. As a result, obtaining a pool of geotagged tweets that expressed a moral stance regarding stay at home orders was difficult, and yielded a much smaller dataset than what was originally expected at the start of the design process.  
	\item Collaboration issues are exacerbated by tight project timelines. We found that in the early stages our collaborators were unable to articulate what they wanted out of the designs: their earliest requirements were a basic COVID-19 dashboard with political associations overlayed on the map. These findings were similar to those described by Sondag et al.~\cite{sondag2022visual}. However, we found that true requirements were more easily assessed by analyzing what points were discussed during meetings and what researchers were most focused on in their analysis. Collaborators also had insufficiently defined goals, and we found that there was not enough time to allow proper collaboration practices to mature (for example, a social science collaborator published project joint results without crediting the whole team). If time allows it, we would recommend relying on "ethnographic" methods rather than interviews for requirement gathering. We would also recommend repeatedly revisiting and enforcing collaborative practices during group meetings.
	\item Keep designs generalizable. Because our study changed so rapidly, by the time the data collection and design process was finished, the original topic was considered "obsolete", with different topics taking over the online debate. In addition, our original data came from Twitter, which now has stricter limits on the amount of data scraping that can be used. In turn, our work needs to adapt to different topics, as well as different data sources to stay relevant. Our end design generalized well across case studies and datasets.
\item Dissemination issues stemming from data. Disappointingly, during the publication process we met resistance due to a misalignment of review expectation and design practice that couldn't be rectified within the project timeline. Most of these issues resulted from the unexpected difficulty in gathering a dataset of sufficient size and quality, largely due to the fact that automatic annotation proved to be ineffective, and the project timeline didn't provide sufficient time to manually collect an annotated sufficiently sized dataset. This was a common review issue in both CI/NLP-centered and vis-centric venues. As a result, several reviewers either claimed we had an insufficient dataset when using manual data, or we had an insufficient algorithm when using automatic annotation. 
\end{enumerate}

\section{Conclusion}
In conclusion, our project was successful in producing insights,  and thus in achieving its proposed goals. It was also satisfying to be able to contribute to a better understanding of the pandemic policies, in particular at a time when vaccines were not yet available. At the same time, it was a difficult, extremely intensive experience, leading to several cases of burnout and dissatisfaction on the team, which were then aggravated by repeated difficulties in getting the methods published. If at all possible, we would rather embark on projects where (most of) the data has been already collected, and (most of) the requirements can be safely established during requirements engineering.

\section*{Acknowledgments}
We thank Juan Trelles and our other colleagues at the Electronic Visualization Laboratory for their technical and emotional support. This work was partially supported by awards 
from the U.S. National Science Foundation (IIS-2031095, CNS-1828265, CDSE-1854815) and the U.S. National Institutes of Health (NLM R01LM012527, NCI R01CA258827).

\bibliographystyle{abbrv-doi}

\bibliography{template}
\end{document}